\documentclass[preprint,preprintnumbers,nofootinbib,tightenlines,12pt]{revtex4}

\usepackage{graphicx}
\usepackage{amssymb}
\usepackage{amsmath,mathrsfs,verbatim}
\usepackage{times}
\usepackage{latexsym}
\def\beq{\begin{equation}}
\def\eeq{\end{equation}}
\def\bey{\begin{eqnarray}}
\def\eey{\end{eqnarray}}

\def\lsim{\mathrel{\raise.3ex\hbox{$<$\kern-.75em\lower1ex\hbox{$\sim$}}}}
\def\gsim{\mathrel{\raise.3ex\hbox{$>$\kern-.75em\lower1ex\hbox{$\sim$}}}}

\newcommand{\be}{\begin{equation}}
\newcommand{\ee}{\end{equation}}

\newcommand{\mx}{m_X}
\newcommand{\ax}{\alpha_X}
\newcommand{\mphi}{m_\phi}

\begin{document}

\title{Direct Detection Portals for Self-interacting Dark Matter}

\author{Manoj Kaplinghat$^{a}$, Sean Tulin$^{b}$, and Hai-Bo Yu$^{c}$
\vspace{5mm}
\\
$^{a}$  \normalsize\emph{Department of Physics and Astronomy, University of California, Irvine, California 92697, USA}  \vspace{1mm} \\ 
$^{b}$  \normalsize\emph{Michigan Center for Theoretical Physics, University of Michigan, Ann Arbor, MI 48109, USA}  \vspace{1mm} \\ 
$^{c}$  \normalsize\emph{Department of Physics and Astronomy, University of California, Riverside, CA, 92507, USA}
}

\date{\today}

\preprint{MCTP-13-33, CETUP2013-018}

\begin{abstract}
\vspace*{.0in}

Dark matter self-interactions can affect the small scale structure of the Universe, reducing the central densities of dwarfs and low surface brightness galaxies in accord with observations.  From a particle physics point of view, this points toward the existence of a $1 - 100$ MeV particle in the dark sector that mediates self-interactions.  Since mediator particles will generically couple to the Standard Model, direct detection experiments provide sensitive probes of self-interacting dark matter.  We consider three minimal mechanisms for coupling the dark and visible sectors: photon kinetic mixing, $Z$ boson mass mixing, and the Higgs portal.  Self-interacting dark matter motivates a new benchmark paradigm for direct detection via momentum-dependent interactions, and ton-scale experiments will cover astrophysically motivated parameter regimes that are unconstrained by current limits.  Direct detection is a complementary avenue to constrain velocity-dependent self-interactions that evade astrophysical bounds from larger scales, such as those from the Bullet Cluster.  

\end{abstract}


\maketitle

\section{Introduction}

Dark and visible matter have very different distributions in the Universe: dark matter (DM) forms diffuse halos (e.g., observed via graviational lensing maps), while visible matter undergoes dissipative dynamics and tends to clump into galaxies and stars.   However, this does not preclude the possibility of new dark sector interactions beyond the usual collisionless DM paradigm.  DM could have a large cross section for scattering with other DM particles and this scenario, dubbed self-interacting DM (SIDM)~\cite{1992ApJ...398...43C,Spergel:1999mh}, can affect the internal structure (mass profile and shape) of DM halos compared to collisionless DM.  In turn, astrophysical observations of structure, compared to numerical N-body simulations, can probe the self-interacting nature of DM.\footnote{Here we assume a single DM component with {\it nondissipative} self-interactions.  It is possible for a subdominant fraction of DM to interact via dissipative processes \cite{Fan:2013tia}.}  It is worth emphasizing that tests of self-interactions can shed light on the nature of DM {\it even if DM is completely decoupled with respect to traditional DM searches}.  

In fact, there are long-standing issues on small scales that may point toward SIDM.  Dwarf galaxies are natural DM laboratories since in these galaxies DM tends to dominate baryons inside the optical radius. Observations indicate that the central regions of well-resolved dwarf galaxies exhibit cored profiles~\cite{Moore:1994yx,Flores:1994gz}, as opposed to steeper cusp profiles found in collisionless DM-only simulations~\cite{Navarro:1996gj}.  Cored profiles have been inferred in a variety of dwarf halos, including within the Milky Way (MW)~\cite{Walker:2011zu}, other nearby dwarfs~\cite{2011AJ....141..193O} and low surface brightness galaxies \cite{2008ApJ...676..920K}.  An additional problem concerns the number of massive dwarf spheroidals in the MW.  Collisionless DM simulations have a population of subhalos in MW-like halos that are too massive to host any of the known dwarf spheriodals but whose star formation should not have been suppressed by ultraviolet feedback~\cite{BoylanKolchin:2011dk}.  While these apparent anomalies are not yet conclusive -- e.g., baryonic feedback effects may be important~\cite{Governato:2012fa,Brooks:2012vi} -- recent state-of-the-art SIDM N-body simulations have shown that self-interactions can modify the properties of dwarf halos to be in accord with observations, without spoiling the success of collisionless DM on larger scales and being consistent with halo shape and Bullet Cluster bounds~\cite{Vogelsberger:2012ku,Rocha:2012jg,Peter:2012jh,Zavala:2012us}.  

The figure of merit for DM self-interactions is cross section per unit DM mass, $\sigma/m_X$, where $X$ is the DM particle.  To have an observable effect on DM halos over cosmological timescales, the cross section per unit mass must be of order
\be
{\sigma}/{m_\chi} \sim 1 \; {\rm cm^2/g} \; \approx \; 2 \times 10^{-24} \; {\rm cm^2/GeV} \, 
\ee
or larger. 
From a particle physics perspective, this value is many orders of magnitude larger than expected from weak-scale physics.  For a typical weakly-interacting massive particle (WIMP), the cross section is $\sigma \! \sim\!10^{-36} \; {\rm cm^2}$ and the mass is $m_X \sim 100$ GeV, giving $\sigma/m_X \sim 10^{-38} \; {\rm cm^2/GeV}$.  Evidence for self-interactions would therefore point toward a new dark mediator particle $\phi$ that is much lighter than the weak scale, typically $m_\phi \sim 1 -100$ MeV~\cite{Tulin:2013teo}.  Light mediators have been widely studied within many different contexts, e.g., hidden sector models~\cite{Feng:2008mu}, light DM models~\cite{Boehm:2003hm,Lin:2011gj}, Sommerfeld-enhanced models for indirect detection signals~\cite{ArkaniHamed:2008qn,Pospelov:2008jd}, and in connection with supersymmetry and the WIMP miracle~\cite{Feng:2008ya,Cheung:2009qd,Katz:2009qq,Morrissey:2009ur}. The light force carrier can be probed by low energy experiments with high luminosities~\cite{Essig:2009nc,Bjorken:2009mm, Batell:2009di,AmelinoCamelia:2010me,Merkel:2011ze,Abrahamyan:2011gv,Balewski:2013oza,Izaguirre:2013uxa,Essig:2013vha}. The effect of self-interactions can also be searched through various other astrophysical observations {e.g.}~\cite{Zentner:2009is,CyrRacine:2012fz,Aarssen:2012fx,Dawson:2012fx,Kahlhoefer:2013dca,Linder:2013dga,Harvey:2013tfa,Kouvaris:2011gb,Fan:2013yva,Fan:2013tia,Bramante:2013nma,Pearce:2013ola,Goldman:2013qla,Ma:2013yga,Cyr-Racine:2013fsa}.

Astrophysical tests of self-interactions can probe DM that is decoupled from the Standard Model (SM).  However, the dark and visible sectors are likely not to be {\it completely} decoupled.  $\phi$ particles in the early Universe must decay, otherwise they over-produce dominate DM. Unless additional states are introduced, $\phi$ must decay to SM particles.  These decay products --- typically electrons, positrons, photons, and neutrinos --- must not spoil the light element abundances predicted by Big Bang Nucleosynthesis (BBN) and or produce too much entropy.  This requires a $\phi$ lifetime less than $\sim 1$ second, although this can be relaxed slightly as we discuss later.

\begin{figure}
\includegraphics[scale=1]{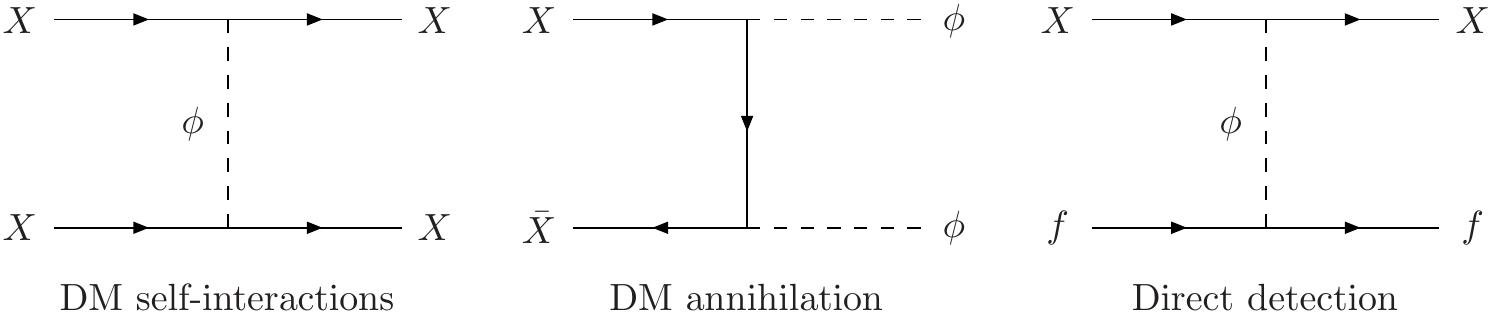}
\caption{Feynman diagrams arising from DM particle $X$ coupled to a dark force mediator $\phi$.  Self-interactions modify DM halos, while annihilation can give the observed DM relic density.  Direct detection experiments are highly sensitive to potential couplings of $\phi$ to SM fermions $f$, which allow decays $\phi \to f \bar f$ before BBN.}
\label{feynman}
\end{figure}

In this paper, we study the reach for DM direct detection experiments to probe SIDM.  We consider different scenarios for how $\phi$ might couple to SM fermions by mixing with the photon, $Z$ or Higgs bosons, depending on the spin of $\phi$.  Since BBN constraints give a {\it lower} bound on $\phi$ couplings to the SM, this defines a {\it minimum} value for the direct detection cross section.  As we show, the predicted SIDM direct detection range lies within the sensitivity reach of upcoming experiments such as  XENON1T~\cite{Aprile:2012zx}, LUX~\cite{Akerib:2012ys}, and SuperCDMS~\cite{Brink:2012zza}, although the quantitative details are model dependent.  The essential physics is summarized in Fig.~\ref{feynman}.  Both self-interactions and direct detection arise through the same light mediator $\phi$, which can also provide an annihilation channel for setting the DM relic density.

Direct detection searches for SIDM are highly complementary to other astrophysical probes.  For nuclear recoils, direct detection experiments are sensitive to DM masses $m_X \gtrsim {10~\rm GeV}$ due to threshold limitations.  In this mass range, the self-interaction cross section tends to be more velocity-dependent, becoming suppressed at higher velocity (like Rutherford scattering)~\cite{Tulin:2013teo}.  In this case, DM can become effectively collisionless within larger DM halos, which have larger characteristic velocities.  Tests for self-interactions within large DM halos, such as in merging clusters~\cite{Randall:2007ph,Dawson:2011kf} or halo shape observables~\cite{Feng:2009hw}, are most sensitive to DM masses $m_X \lesssim {\rm 10\;  GeV}$ where the self-interaction cross section tends to be more velocity-independent.  We also note that direct detection experiments can also probe low mass SIDM as well, via electron recoils~\cite{Essig:2011nj,Essig:2012yx}, although the effective reach in couplings is much reduced compared to nuclear recoils.  

In the remainder of this work, we first present simple particle physics models for SIDM in Sec.~\ref{sec:model}.  Then we discuss different portals for how the dark and visible sectors may be coupled in Sec.~\ref{sec:portal}.  We present our results in Sec.~\ref{sec:results}, showing how direct detection and other constraints map onto SIDM parameter space.  Lastly, we conclude in Sec.~\ref{sec:conclusions}.


\section{Particle physics of self-interacting dark matter}
\label{sec:model}

As a minimal model for SIDM, we consider DM $X$ to be a Dirac fermion which is coupled to a mediator particle $\phi$.  $X$ is assumed to be a SM gauge singlet with no direct coupling to SM particles.  We take $\phi$ to be either a real vector or scalar, with an interaction given by
\beq
\mathscr{L}_{\rm int} = \left\{ \begin{array}{ll} g_X \bar X \gamma^\mu X \phi_\mu & {\rm vector \; mediator}  \\
 g_X \bar X X \phi & {\rm scalar \; mediator} \end{array} \right.
\eeq
where $g_X$ is a coupling constant.  That is, the mediator $\phi$ is the dark sector counterpart of force particles in the SM, analogous to a $Z$ or Higgs boson. 

Nonrelativistic DM self-scattering, shown in Fig.~\ref{feynman}, can be described a Yukawa potential~\cite{Feng:2009hw,Loeb:2010gj,Buckley:2009in}
\beq
V(r) = \pm \frac{\alpha_X}{r} \, e^{- m_\phi r} \; , 
\eeq
with $+$ ($-$) sign for an repulsive (attractive) interaction.  For the scalar $\phi$ case, DM scattering is purely attractive.  For the vector $\phi$ case, $XX$ or $\bar X\bar X$ scattering is repulsive, while $X \bar X$ is attractive.  Other types of interactions give rise to more complicated potentials~\cite{Bellazzini:2013foa,Liu:2013vha}, which we do not consider.  The differential scattering cross section $d\sigma/d\Omega$ can be computed numerically through a standard partial wave analysis by solving the Schr\"{o}dinger equation with $V(r)$ for the DM two-body wavefunction; see Ref.~\cite{Tulin:2013teo} for details.  This method takes into account nonperturbative effects via multiple scatterings, analogous the the Sommerfeld enhancement for annihilation, which become important when the Born approximation breaks down for $\alpha_X m_X /m_\phi \gtrsim 1$.  In the nonperturbative regime, DM self-scattering has a rich structure, including possible quantum mechanical resonances through bound state formation~\cite{Tulin:2012wi}. 

To compare with results from numerical N-body simulations of SIDM, we consider $\langle \sigma_T \rangle$ as a suitable proxy for the self-interaction cross section $\sigma$ entering into halo observables.  Here, the transfer cross section is defined by $\sigma_T \equiv \int d\Omega \, (1- \cos\theta) \, d \sigma/ d \Omega$, where $(1-\cos\theta)$ is the fractional longitudinal momentum transfer for scattering angle $\theta$.  Moreover, we average $\sigma_T$ over the initial DM velocities $\vec{v}_{1,2}$ by
\be
\langle \sigma_T \rangle = \int \frac{ d^3 v_1 d^3 v_2}{(\pi v_0^2)^{3}} \, e^{-v_1^2/v_0^2}  \, e^{-v_2^2/v_0^2} \, \sigma_T(|\vec{v}_1 -\vec{v}_2|) = \int \frac{d^3 v}{(2\pi v_0^2)^{3/2}} \, e^{-\frac{1}{2} v^2/v_0^2} \, \sigma_T(v) \; ,
\ee
where $v_0$ is the most probable velocity for a given halo and $v = |\vec{v}_1 -\vec{v}_2|$ is the relative velocity.  We choose $v_0$ to be characteristic of different size halos, with larger halos having bigger $v_0$.  The point is that $d \sigma/d \Omega$ has a complicated dependence on both $v$ and $\theta$ in general, and by considering $\langle \sigma_T \rangle$, we are averaging over these dependencies in a physically meaningful way to obtain a single effective self-interaction cross section $\sigma$ for a given halo.\footnote{In principle, since numerical simulations follow the trajectories and velocities of each ``particle,'' these dependencies can be straight-forwardly accounted for.  However, SIDM simulations thus far have been restricted to cross sections with a $v$-dependence that is either constant~\cite{Rocha:2012jg,Peter:2012jh} or motivated within the classical limit ($m_X v/m_\phi \gg 1$)~\cite{Vogelsberger:2012ku,Zavala:2012us} and with isotropic angular dependence.}

Aside from self-interactions, the mediator $\phi$ can play an important role in setting the DM relic density in the early Universe via
$X \bar X \to \phi \phi$, shown in Fig.~\ref{feynman}.  The annihilation cross section is (at the Born level)
\beq \label{anncross}
(\sigma v)_{\rm ann} \approx \frac{\pi \alpha_X^2}{m_X^2}  \times \left\{ \begin{array}{cc} 1 & {\rm vector \; mediator} \\   3v^2/8 & {\rm scalar \; mediator} \end{array} \right. \; 
\eeq
at lowest order in $v$ and $m_\phi$.
Annihilation is $s$-wave for a vector mediator and $p$-wave for a scalar mediator.  For symmetric DM, where $X$ and $\bar X$ are populated equally in the early Universe, the relic density is determined by standard freeze-out, requiring a thermally-averaged cross section $\langle \sigma v \rangle_{\rm ann} \approx 6 \times 10^{-26} \; {\rm cm^3/s}$.  Using Eq.~\eqref{anncross}, the DM relic density is 
\beq \label{Omegadm}
\Omega_{\rm dm} \sim 0.2 \times \left( \frac{6 \times 10^{-26} \; {\rm cm^3/s}}{\langle \sigma v \rangle_{\rm ann}} \right) \sim
0.2 \times \left( \frac{\alpha_X}{10^{-2}} \right)^{-2} \times \left\{ \begin{array}{ll} (m_X/300 \; {\rm GeV})^2 & {\rm vector} \\ (m_X/100 \; {\rm GeV})^2 & {\rm scalar} \end{array} \right.\; ,
\eeq
where we have set $\langle v^2 \rangle = 6 T/m_X \sim 0.3$ for $p$-wave annihilation.  Eq.~\eqref{Omegadm} demonstrates that SIDM keeps the virtues of WIMP miracle intact, with weak-scale mass $m_X$ and coupling $\alpha_X$ giving the observed value $\Omega_{\rm dm} \sim 0.2$.  DM masses below the weak scale are also allowed provided the ratio $\alpha_X/m_X$ remains fixed~\cite{Feng:2008ya}. 

Alternately, DM may be asymmetric, such that $X$ and $\bar X$ are populated unequally in the early Universe due to a primordial $X$-number asymmetry~\cite{Davoudiasl:2012uw,Petraki:2013wwa,Zurek:2013wia}.  In this case, annihilation still plays a crucial role for depleting the symmetric density of $X$ and $\bar X$, leaving behind the residual asymmetric component composed of $X$ only.  The annihilation cross section is required to be {\it larger} than for thermal freeze-out, $\langle \sigma v\rangle_{\rm ann} > 6 \times 10^{-26} \; {\rm cm^3/s}$.  Conversely, for fixed $\alpha_X$, this gives an upper bound on $m_X$ above which annihilation is insufficient to avoid overclosing the Universe. Since the mediator mass is light, the Born annihilation cross section is modified by a Sommerfeld enhancement.  We include the Sommerfeld enhancement in our calculation of the DM relic density~\cite{Feng:2010zp}. Sommerfeld enhancement provides an $\mathcal{O}(1)$ correction to the ratio $\alpha_X/m_X$ required by relic density for $m_X \gtrsim 100 \; {\rm GeV}$, but has virtually no impact for smaller $m_X$ (see Ref.~\cite{Tulin:2013teo}).


\section{Cosmology of Light Mediators}

The presence of the light mediator $\phi$ in the early Universe has important consequences. If it is stable, its comoving number density is  
\begin{eqnarray}
Y_\phi\simeq0.208\frac{g_\phi\xi^3}{g_{*s}}\sim7\times10^{-2}\xi^3,
\label{eq:yphi}
\end{eqnarray}
where we set the internal degrees of freedom of the mediator $g_\phi=3$, the entropy degrees of freedom $g_{*s}=10$, and $\xi$ is the temperature ratio of the dark sector to the SM. Compared to the baryon number density $Y_B\sim10^{-10}$, we see that $\phi$ can easily dominate the energy of the Universe if it is stable. One option to reduce the $\phi$ energy density is to make the dark sector very cold. From Eq.~\ref{eq:yphi}, one would expect that $Y_\phi$ becomes negligible if $\xi\lesssim10^{-3}(10~{\rm MeV}/\mphi)^{1/3}$. However, $\xi$ can not be arbitrary small because DM particles have to be populated in the hidden sector thermal bath, which leads to a lower bound $\xi\gtrsim10^{-3}\left(g_{*s}1~{{\rm GeV}/\mx}\right)^{1/3}$.

To evade the overclosure problem, we assume that $\phi$ is not stable and decays to SM particles. There are many cosmological constraints on the late decay of the mediator particle. Since the mediator has a mass in the range of $1-100$ MeV, its decay products are electrons, positrons and neutrinos. The energy injection from decay products (if lifetime is larger than about a second) may change the BBN light element abundances in a manner inconsistent with observations. For example, for a 100 GeV mass particle decaying to electromagnetic final states, the BBN constraint on the lifetime is $\lesssim 10^{4}$ s~\cite{Jedamzik:2009uy}. However, it should be noted that for $m_\phi$ below the $^4{\rm He}$ binding energy, the effect on BBN abundances will be restricted to destruction of deuterium and lithium isotopes and a reconsideration of the effect of energy injection in this regime with $1-100$ MeV mass decaying particle is warranted. 

The decays also lead to an increase in the entropy and this change can be estimated as~\cite{Kolb:1990vq}
\begin{eqnarray}
\frac{S_f-S_i}{S_i}\simeq g^{1/4}_*\frac{m_\phi Y_\phi\tau^{1/2}_\phi}{m_{\rm pl}}\sim 0.4\xi^3\left(\frac{\mphi}{10~{\rm MeV}}\right)\left(\frac{\tau_\phi}{1~{\rm s}}\right)^{1/2},
\label{eq:entropy}
\end{eqnarray}
where $S_i$ ($S_f$) is the total entropy before (after) $\phi$ decays, and $\tau_\phi$ is the lifetime of $\phi$. In deriving Eq.~\ref{eq:entropy}, we have assumed radiation dominates the energy density before $\phi$ decays, which is valid if $T_D>0.7~{\rm MeV}~(\mphi/10~{\rm MeV})\xi^3$ with $T_D$ as the temperature when $\phi$ decays. This condition can be satisfied in our model parameter space.

If the decay occurs after BBN with $\tau_\phi\gtrsim1$ s, entropy production has to be less than $\sim10\%$ (assuming no change to light element abundances due to it) because the precision measurements of the baryon density through BBN and CMB observations match very well~\cite{Steigman:2012ve}. Therefore, the absence of significant entropy production after BBN will put a strong bound on the lifetime $\tau_\phi$ and the temperature ratio $\xi$. If the mediator decays before BBN, entropy production constraints are non-existent.

We do not undertake a detailed study of the BBN and entropy production constraints but simply note that $\tau_\phi=1$ s is safe from such constraints if $\xi\sim0.3-1$ depending on the mediator mass. We therefore adopt this value of $\tau_\phi$ as a benchmark to set the coupling between $\phi$ and the SM particles, and estimate the direct detection cross section based on it. A smaller value of $\tau_\phi$ (at fixed $\xi$) will lead to a stronger signal in direct detection experiments.

For the simplicity, we will present our results based on $\xi=1$, which can be extended to other cases by  simple rescaling. Since the DM annihilation cross section scales as $\xi$ to obtain the correct relic density~\cite{Feng:2008mu}, $\ax$ is proportional to $\xi^{1/2}$ at fixed $m_X$. This implies that the lower bound on the symmetric SIDM mass from the CMB weakens by a factor of $\xi$. To keep the same self-scattering cross section, the mediator mass scales as $\xi^{1/4}$, $\xi^{1/2}$ and $\xi^{0.1}$ in the Born, resonant and strongly-coupled classical regimes, respectively. Therefore, the direct detection cross section in the limit of zero momentum transfer ($q^2=0$) scales as $\xi^{-1/2}$, $\xi^{-3/2}$ and $\xi^{0.1}$ in the Born, resonant and classical regimes, respectively. For momentum transfer $q \gg m_\phi$ (see Eq.~\eqref{formfactor}), the direct detection cross section scales as $\xi^{1/2}$.  Since the favored $\xi$ is in the range of $0.3-1$, these are all ${\cal O}(1)$ corrections.




\section{Portals for light mediators}
\label{sec:portal}

In this section, we discuss three minimal ways that the dark mediator $\phi$ may be coupled to the SM.  Depending on its spin, $\phi$ can mix with photon or $Z$ boson if it is a vector, or it can mix with the Higgs boson if it is a scalar.  This mixing allows $\phi$ particles, produced thermally in the early Universe, to decay.  That these decays do not affect BBN and dilute the baryon density puts a constraint on the mixing parameter, which in turn provides a lower bound on the DM direct detection cross section.

\subsection{Vector mediator case: photon and $Z$ mixing}

Since gauge symmetries exist in the visible sector, it is natural to speculate that they may be present in the dark sector as well.  The simplest example is a $U(1)_X$ symmetry under which DM is charged, with $\phi$ being the corresponding vector boson.  We assume that the $U(1)_X$ is spontaneously broken, say by a dark Higgs, to generate a mass $m_\phi$.  (The massless case has also been discussed in Refs.~\cite{Ackerman:2008gi,Feng:2009mn}.)

If there exist particles that are charged under both the $U(1)_X$ and the SM $SU(2)_L \times U(1)_Y$ gauge symmetries, this leads to mixing between $\phi$ and the photon $\gamma$ or $Z$ boson.  The most relevant interactions are
\beq \label{Lmix1}
\mathscr{L}_{\rm mixing} = \frac{\varepsilon_\gamma}{2}\,  \phi_{\mu \nu}  F^{\mu\nu} + \delta m^2 \, \phi_\mu Z^{\mu} 
\eeq
where $\phi_{\mu\nu} \equiv \partial_\mu \phi_\nu - \partial_\nu \phi_\mu$ is the $U(1)_X$ field strength.  The first term corresponds to photon kinetic mixing~\cite{Holdom:1985ag}, which has been widely studied as a DM portal~\cite{Foot:2004pa,Feldman:2006wd,Pospelov:2008jd, ArkaniHamed:2008qn} and in dark photon searches~\cite{Bjorken:2009mm}.  The second term corresponds to mass mixing with the $Z$~\cite{Babu:1997st,Davoudiasl:2012ag}.  This term explicitly breaks gauge invariance, but can arise via higher-dimensional operators with additional Higgs and dark Higgs insertions.   Defining $\varepsilon_Z \equiv \delta m^2/m_Z^2$, we work in the limit of small mixing parameters, $\varepsilon_{\gamma,Z} \ll 1$.

At $\mathcal{O}(\varepsilon_{\gamma,Z})$, mixing via Eq.~\eqref{Lmix1} induces a coupling of $\phi$ to SM fermions, given by
\beq \label{LDD1}
\mathscr{L}_{\rm int} = \Big(  \varepsilon_\gamma e J_{\rm em}^\mu + \varepsilon_Z \frac{g_2}{c_W}  J_{\rm NC}^\mu \Big) \phi_\mu \, ,
\eeq
where the electromagnetic and weak neutral currents are, respectively,
\beq
J_{\rm em}^\mu = \sum_f Q_f \bar f \gamma^\mu f \, , \quad J_{\rm NC}^\mu = \sum_f  \bar f \gamma^\mu (T_{3f} P_L - Q_f s_W^2 ) f  \, ,
\eeq
Here, $g_2$ is the $SU(2)_L$ coupling, $\theta_W$ is the weak mixing angle (with $s_W \equiv \sin\theta_W$ and $c_W \equiv \cos\theta_W$), and $Q_f$ and $T_{3f}$ denote the charge (in units of $e$) and weak isospin for SM fermion $f$.

The decay rate for $\phi$ can be straightforwardly computed from Eq.~\eqref{LDD1}.  Since $m_\phi \sim 1 - 100$ MeV is typically preferred by solving small scale structure anomalies with SIDM, only $e^+ e^-$, $\nu \bar \nu$, and photon final states are allowed.  For kinetic mixing, decay is dominated by $\phi \to e^+ e^-$, with decay rate and lifetime
\beq
\Gamma_\phi =  \frac{\alpha_{\rm em} m_\phi \varepsilon_\gamma^2 }{3} \quad \Rightarrow \quad  \tau_\phi \approx {\rm 3 \; seconds} \times \left( \frac{\varepsilon_\gamma}{10^{-10}} \right)^{-2} \left( \frac{m_\phi}{\rm 10 \; MeV} \right)^{-1} \; ,
\eeq
where $\alpha_{\rm em}$ is the electromagnetic fine structure constant.
For $Z$ mixing, neutrino final states dominate since there are three families and also the electron's weak vector charge is suppressed.  The total decay rate and lifetime are
\beq
\Gamma_\phi = \frac{\alpha_{\rm em} m_\phi \varepsilon_Z^2 (1-s_W^2 + 2s_W^4)}{6 s_W^2 c_W^2} \quad \Rightarrow \quad  \tau_\phi \approx {\rm 1 \; second} \times \left( \frac{\varepsilon_Z}{10^{-10}} \right)^{-2} \left( \frac{m_\phi}{\rm 10 \; MeV} \right)^{-1} \; ,
\eeq
and the branching ratios are ${\rm BR}(\phi \to \nu \bar \nu) \approx 6/7$ and ${\rm BR}(\phi \to e^+ e^-) \approx 1/7$.  Ensuring that decays $\phi \to e^+ e^-$ do not occur after BBN requires a lifetime $1/\Gamma_\phi \lesssim {\rm 1 \; second}$.  Therefore, the mixing parameters are constrained to be $\epsilon_{\gamma,Z} \gtrsim 10^{-10} \times \sqrt{{\rm 10 \; MeV}/m_\phi}$~\cite{Lin:2011gj}.

The upper bound from the low energy beam dump experiments is $\varepsilon_\gamma \lesssim10^{-7}$ for $\mphi\lesssim400~{\rm MeV}$~\cite{Bjorken:2009mm}, while the region $10^{-10} \lesssim \varepsilon_\gamma \lesssim 10^{-7}$ is excluded for $\mphi\lesssim100~{\rm MeV}$ by energy loss arguments in supernovae~\cite{Dent:2012mx} (although this constraint depends sensitively on assumptions about the temperature and size of the supernova core).

Next, we discuss direct detection.  Given Eq.~\eqref{LDD1}, direct detection signals arise from DM scattering on nuclei via $\phi$ exchange, shown in Fig.~\ref{feynman}.  In general, we can parametrize the coupling of $\phi$ to protons ($p$) and neutrons ($n$) as 
\beq \label{nucleoncouplings}
\mathscr{L}_{\rm int} = e \phi_\mu \big( \varepsilon_p \bar p \gamma^\mu p + \varepsilon_n \bar n \gamma^\mu n \big) \; ,
\eeq
where $\varepsilon_{p,n}$ are the effective nucleon couplings (in units of $e$).  For kinetic mixing or $Z$ mixing, they are given by
\beq
\varepsilon_p = \varepsilon_\gamma + \frac{\varepsilon_Z}{4 s_W c_W} (1-4 s_W^2) \approx \varepsilon_\gamma + 0.05 \varepsilon_Z \, , \quad \varepsilon_n = - \frac{\varepsilon_Z}{4 s_W c_W} \approx - 0.6 \varepsilon_Z \, .
\eeq
That is, kinetic mixing couples $\phi$ to protons only, since only protons carry electric charge, while $Z$ mixing couples $\phi$ mainly to neutrons, since the weak charge of the proton is $(1-4 s_W^2) \ll 1$.  For arbitrary $\varepsilon_{\gamma,Z}$, this scenario is maximally isospin-violating in the sense that any ratio of $\varepsilon_p/\varepsilon_n$ is allowed~\cite{Frandsen:2011cg}.
For a given nucleus $N$ with proton number $Z$ and mass number $A$, the spin independent cross section is, in the limit of zero momentum transfer ($q^2=0$),
\beq \label{sigmaSI}
\sigma_{X N}^{\rm SI} = \frac{16 \pi \alpha_{\rm em} \alpha_X \mu_{X N}^2 }{m_\phi^4} \,\big( \varepsilon_p Z  + \varepsilon_n (A-Z) \big)^2 \; ,
\eeq
where $\mu_{XN} = m_X m_N/(m_X + m_N)$ is the DM-nucleus reduced mass.  

Cosmological constraints imply that the SIDM direct detection cross section may be within the reach of present or future searches.  For kinetic mixing, the SI cross section on protons is
\beq
\sigma_{Xp}^{\rm SI} \approx 1.5 \times 10^{-24} \; {\rm cm^2} \times \varepsilon_\gamma^2 \times \left( \frac{\alpha_X}{10^{-2}} \right) \left( \frac{m_\phi}{\rm 30 \; MeV}\right)^{-4} \; ,
\eeq
while for $Z$ mixing, the SI cross section on neutrons is
\beq
\sigma_{Xn}^{\rm SI} \approx 5 \times 10^{-25} \; {\rm cm^2} \times \varepsilon_Z^2 \times \left( \frac{\alpha_X}{10^{-2}} \right) \left( \frac{m_\phi}{\rm 30 \; MeV}\right)^{-4} \; .
\eeq
If we take $\varepsilon_{\gamma,Z} \gtrsim 10^{-10}$ as a lower bound from BBN, the predicted SI cross section per nucleon is predicted to be greater than $\sim 10^{-45} - 10^{-44} \; {\rm cm^2}$, which is in the reach of direct detection experiments. 

For SIDM, the typical mediator mass is $m_\phi \sim 1 - 100$ MeV, which is comparable or less than the typical momentum transfer $q \sim 50$ MeV in nuclear recoils.  Therefore, unlike typical WIMPs, nuclear recoil interactions for SIDM are momentum-dependent and cannot be approximated by a contact interaction~\cite{Chang:2009yt,Feldstein:2009tr,Fornengo:2011sz}.  The differential recoil rate per target nucleus is
\bey
\frac{dR}{dE_R} &=& \frac{\rho_{\rm DM}}{m_X} \int_{v_{\rm min}} d^3v \, v \, f(\vec{v}) \, \frac{d\sigma^{\rm SI}_{XN}(v,E_R)}{dE_R} \\
&=& \frac{\rho_{\rm DM}}{m_X} \int_{v_{\rm min}} d^3v \, v \, f(\vec{v}) \, \left(\frac{d\sigma^{\rm SI}_{XN}(v,E_R)}{dE_R} \right)_{q^2=0}  \times  \frac{m_\phi^4}{(m_\phi^2 + q^2)^2} \, , \label{q2rate}
\eey
where $E_R$ is the nuclear recoil energy, $v_{\rm min} = \sqrt{ m_N E_R/ 2 \mu_{XN}^2 }$ is the minimum DM velocity for a given $E_R$, $f({\vec v})$ is the local DM velocity distribution, and $\rho_{\rm DM}$ is the local DM density.  The last term in Eq.~\eqref{q2rate} shows how the momentum transfer, given by $q = \sqrt{2 m_N E_R}$, provides a suppression of the recoil rate compared to taking $q^2=0$ assumed by a contact interaction.  Clearly lighter target nuclei and lower thresholds are favorable for detecting SIDM.

Since cross section limits quoted by the experimental collaborations assume a contact interaction, one must perform a reanalysis of experimental data to derive constraints on SIDM.  Ref.~\cite{Fornengo:2011sz} performed such an analysis for the kinetic mixing model we consider for several light mediator masses ($m_\phi = 0$, $10$, and $30$ MeV), although not within an SIDM context.  In the present work, we account for this suppression effect in a simplified manner by multiplying the total $q^2 = 0$ cross section by a $q^2$-dependent form factor:
\beq \label{formfactor}
\sigma_{XN}^{\rm SI}(q^2) = \sigma_{XN}^{\rm SI}(q^2=0) \times \frac{m_\phi^4}{(m_\phi^2 + q^2)^2} \; .
\eeq
We take a fixed value of $q$, corresponding to a typical momentum transfer for a given experiment.  We assume that the experimentally quoted cross section limits apply directly to Eq.~\eqref{formfactor}, and then we extract what are the constraints on the particle physics parameters.  We defer a complete reanalysis over the whole range of $m_\phi$ for SIDM to future work.  However, we find that this simple prescription is able to reproduce the XENON100 and CDMS reanalysis in Ref.~\cite{Fornengo:2011sz}, for an appropriate choice of $q$, with little dependence on astrophysical halo properties considered therein.

It is also useful to consider the direct detection cross section for DM scattering on electrons, which was proposed as an avenue for detecting sub-GeV DM~\cite{Essig:2011nj,Essig:2012yx}.  The SI cross section is
\beq \label{sigmae}
\sigma_{X e}^{\rm SI} = \frac{16 \pi \alpha_{\rm em} \alpha_X \mu_{X e}^2 \varepsilon_\gamma^2}{m_\phi^4}   
\approx 
4 \times 10^{-29} \; {\rm cm^2} \times \varepsilon_\gamma^2 \times \left( \frac{\alpha_X}{10^{-2}} \right) \left( \frac{m_\phi}{\rm 10 \; MeV}\right)^{-4} \; .
\eeq
Scattering on electrons is mainly sensitive to $\varepsilon_\gamma$, while the $\varepsilon_Z$ term is suppressed by the electron weak charge and we have neglected it in Eq.~\eqref{sigmae}.  Current sensitivities from XENON10 at the level of $\sigma_{Xe}^{\rm SI} \sim 10^{-38} \; {\rm cm^2}$ for sub-GeV mass DM therefore probe this scenario down to $\varepsilon_\gamma \sim 10^{-4}$~\cite{Essig:2012yx}.  

Lastly, we discuss indirect detection.  If the DM density is symmetric, i.e., populated by both $X$ and $\bar X$, then $X \bar X \to \phi \phi$ annihilation, with $\phi$ decaying to SM states, can provide visible astrophysical signals.  Observations of the ionization history of the Universe from the cosmic microwave background (CMB) constrain DM annihilation during recombination~\cite{Galli:2009zc,Slatyer:2009yq}, in particular the process $X \bar X \to \phi \phi \to e^+ e^- e^+ e^-$.  A recent analysis combining several cosmological datasets found $m_X > 30$ GeV for $\langle \sigma v\rangle_{\rm ann}$ fixed by relic abundance and if ${\rm BR}(\phi \to e^+ e^-) = 1$~\cite{Lopez-Honorez:2013cua,Madhavacheril:2013cna}, as in the case of kinetic mixing.  Implications for indirect detection searches (e.g., Fermi and AMS-2) are also important and will be considered elsewhere~\cite{KLY}.  On the other hand, in the case of $Z$ mixing, these limits are weakened by a factor of $\sim 7$ since $\phi$ predominantly decays to neutrinos.  In this scenario, DM capture and annihilation in the sun may provide neutrino-rich signals that may be observed by IceCube. Indirect detection signals from DM annihilation are not present if DM is asymmetric.

\subsection{Scalar mediator case: Higgs mixing}

If the dark mediator $\phi$ is a scalar, the leading renormalizable couplings to SM particles arise through the Higgs portal~\cite{Patt:2006fw,MarchRussell:2008yu,Ahlers:2008qc,Andreas:2008xy,Arina:2010wv,Chu:2011be,Djouadi:2011aa,Bhattacherjee:2013jca,Greljo:2013wja,Bian:2013wna,Choi:2013qra}.  Assuming $\phi$ is a real scalar singlet, the relevant terms in the scalar potential are
\beq
V(H,\phi) \supset (a \phi + b \phi^2) |H|^2
\eeq
where $H$ is the Higgs doublet and $a,b$ are coupling constants.  After electroweak symmetry breaking, mixing arises between $\phi$ and the physical Higgs boson $h$ due to the Higgs vacuum expectation value (vev) $v\approx 246$ GeV.  In the limit $a,m_\phi \ll v, m_h$, this mixing angle is $\varepsilon_h \approx a v /m_h^2$.  This generates an effective $\phi$ coupling to SM fermions
\beq
\mathscr{L}_{\rm int} = - \frac{m_f \varepsilon_h }{v} \bar f f \phi \, .
\eeq
For $m_\phi \sim 1 - 100$ MeV, the dominant decay channel is $\phi \to e^+ e^-$, while $\phi \to \gamma\gamma$ is highly suppressed.\footnote{For the Higgs boson of mass $m_h \approx 125$ GeV, the branching ratio to $\gamma\gamma$ is much larger than to $e^+ e^-$, with ${\rm BR}(h \to \gamma \gamma) \approx 2 \times 10^{-3}$ and ${\rm BR}(h \to e^+ e^-) \approx 5 \times 10^{-9}$.  However, a similar conclusion does not hold for $\phi$ due to its much smaller mass entering the $\gamma\gamma$ loop amplitude.  The second and third generation fermion loop amplitudes accidentally cancel the $W$ boson loop amplitude to $\sim 1\%$, and therefore the contribution from first generation fermion loops is crucial, including hadronic effects for quarks, which is beyond the scope of this paper.}  The decay rate and lifetime for $\phi$ are
\beq
\Gamma_\phi = \frac{\varepsilon_h^2 m_e^2  m_\phi}{8 \pi v^2} \quad \Rightarrow \quad
\tau_\phi \approx {\rm4 \; seconds} \times \left(\frac{\varepsilon_h}{10^{-5}}\right)^{-2} \left( \frac{m_\phi}{{\rm 10 \; MeV}} \right)^{-1} \; .
\eeq
Therefore, having $\phi$ decay before BBN implies that $\varepsilon_h \gtrsim 10^{-5}$.

Coupling the Higgs to the dark sector can lead to invisible Higgs decays.  The $a$ term allows for $h \to X \bar X$, provided $m_X < m_h/2$, while the $b$ term allows for $h \to \phi \phi$ (with $\phi$ escaping the detector as missing energy before decay).  Assuming an otherwise SM-like Higgs boson, the resulting invisible partial widths are
\beq
\Gamma(h \to X \bar X)  = \frac{\alpha_X \varepsilon_h^2 m_h}{2} \Big(1 - \frac{4m_X^2}{m_h^2}\Big)^{3/2} \, , \quad \Gamma(h \to \phi \phi) = \frac{b^2 v^2}{8\pi m_h}  \, .
\eeq
The latter channel is utterly negligible if we require no fine-tuning in the $\phi$ mass, such that the vev-induced mass, $b v^2$, is smaller than $m_\phi^2$.  If $|b| v^2 \lesssim m_\phi^2$, then ${\rm Br}(h \to \phi \phi) \lesssim 10^{-10} \times (m_\phi/100 \; {\rm MeV})^{4}$.  On the other hand, $h \to X \bar X$ can be phenomenologically relevant for the LHC.  Invisible Higgs branching ratios larger than $10\%$ ($1\%$) can be achieved for $\alpha_X \varepsilon_h^2 \gtrsim 10^{-5}$ ($10^{-6}$).  For $m_X \gtrsim 10 \; {\rm GeV}$, direct detection constraints strongly exclude such values (see below), but for lighter DM masses this remains an open possibility and is one of the few ways that the LHC can probe SIDM.

For direct detection, the SI cross section is given by Eq.~\eqref{sigmaSI} with
\beq
\varepsilon_{p,n} = \frac{m_{p,n} \varepsilon_h }{e v} \Big( 1 - \tfrac{7}{9} f_{TG}^{(p,n)} \Big) \approx 3 \times 10^{-3} \times \varepsilon_h \, ,
\eeq
where we take $f_{TG}^{(p,n)} \approx 0.943$ for the gluon hadronic matrix element~\cite{Bhattacherjee:2013jca}. As opposed to the vector case, SI scattering is isospin-conserving, with equal cross sections for scattering on protons and neutrons, given by
\beq
\sigma_{Xp}^{\rm SI} \approx 2 \times 10^{-29} \; {\rm cm}^2 \times \varepsilon_h^2 \times \left( \frac{\alpha_X}{10^{-2}} \right) \left( \frac{m_\phi}{\rm 30 \; MeV}\right)^{-4} \; .
\eeq
Since decays before BBN requires $\varepsilon_h \gtrsim 10^{-5}$, this scenario is strongly excluded by current direct detection bounds for $m_X \gtrsim 10$ GeV.  As opposed to the vector mediator case, where quark and lepton couplings are universal, the scalar mediator case is comparatively more constrained by direct detection due to the larger couplings to quarks compared to electrons.


\section{SIDM models for direct detection}
\label{sec:results}

\begin{figure}
\includegraphics[scale=0.63]{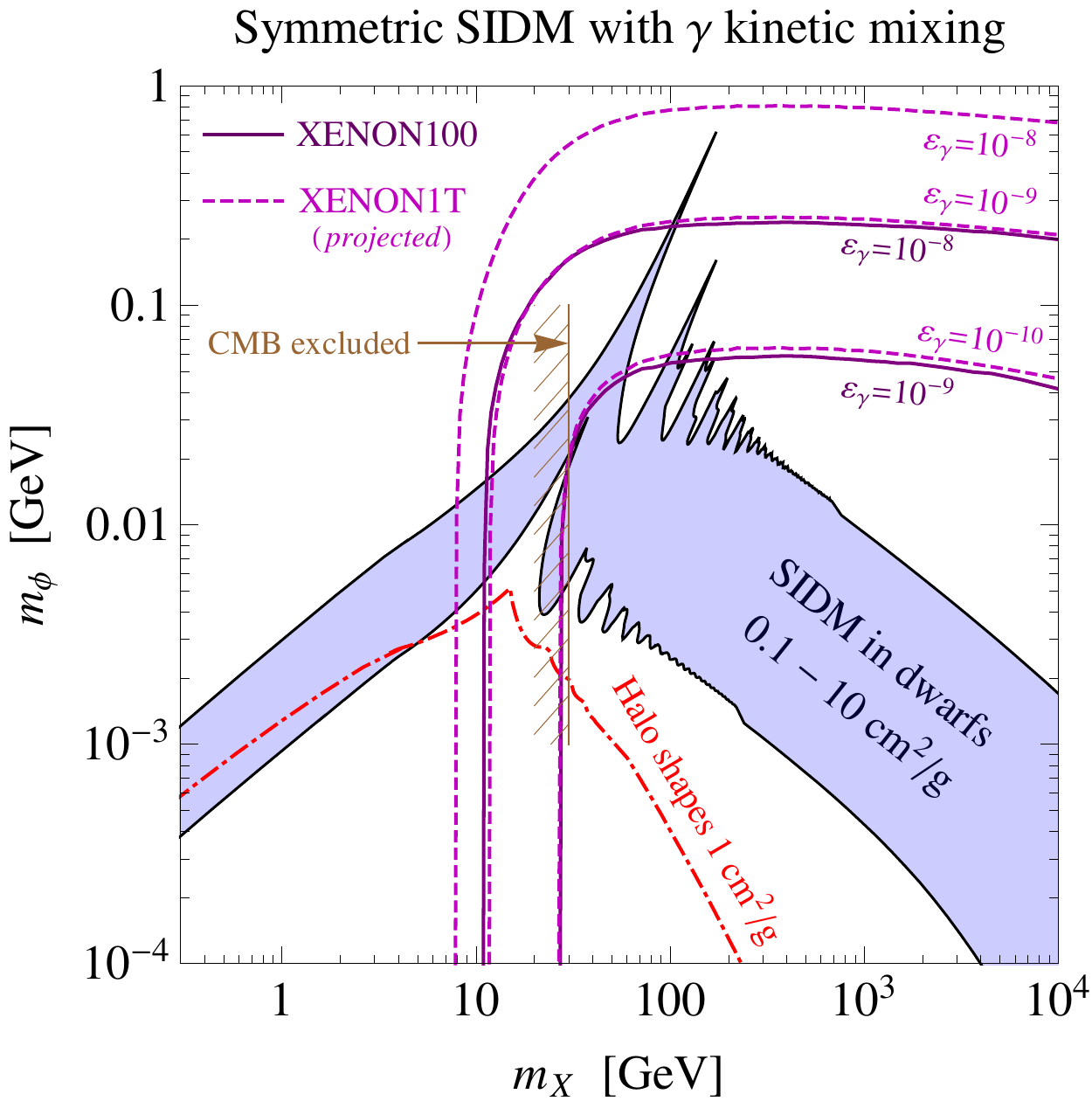} 
\includegraphics[scale=0.63]{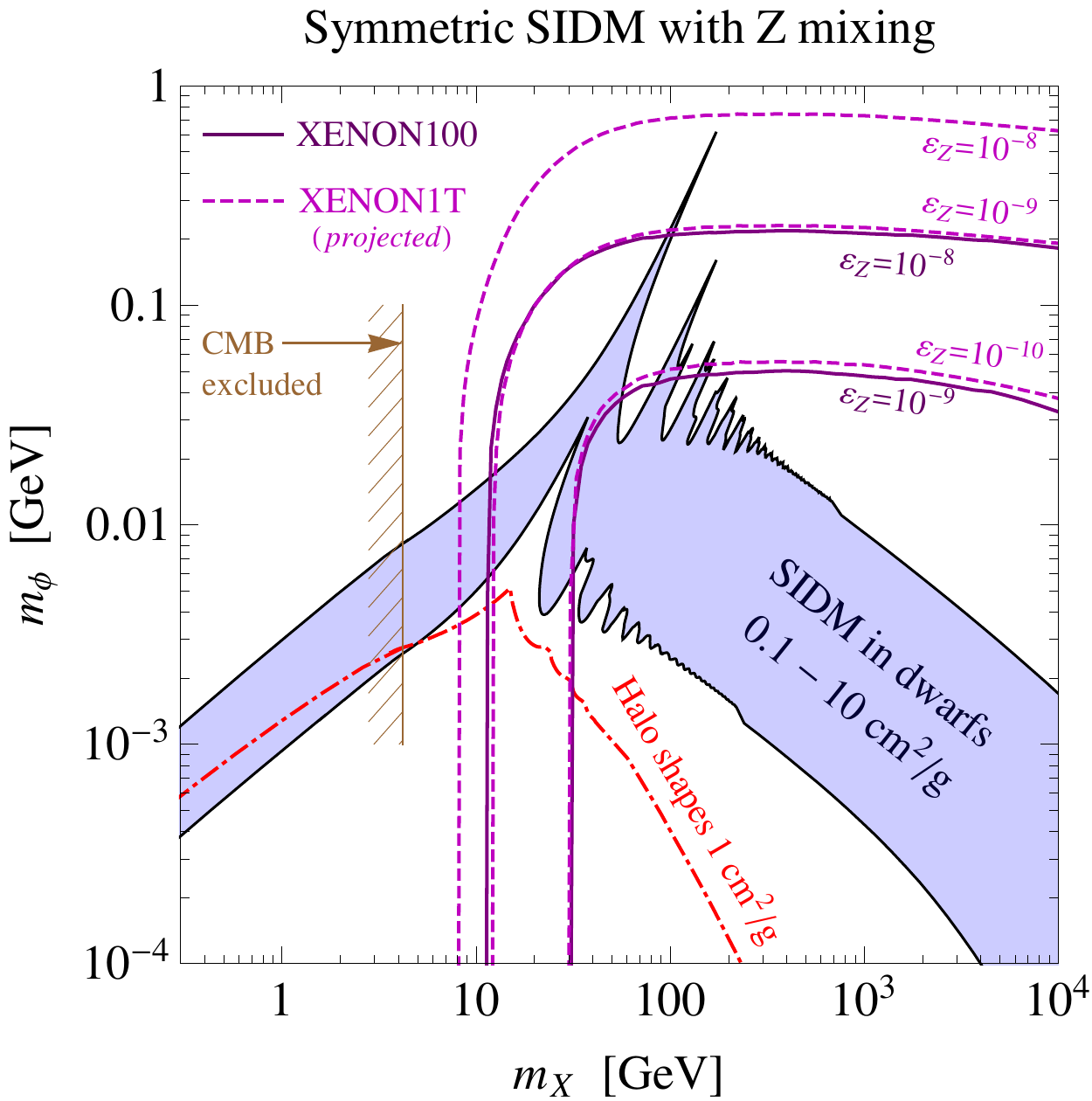}
\includegraphics[scale=0.63]{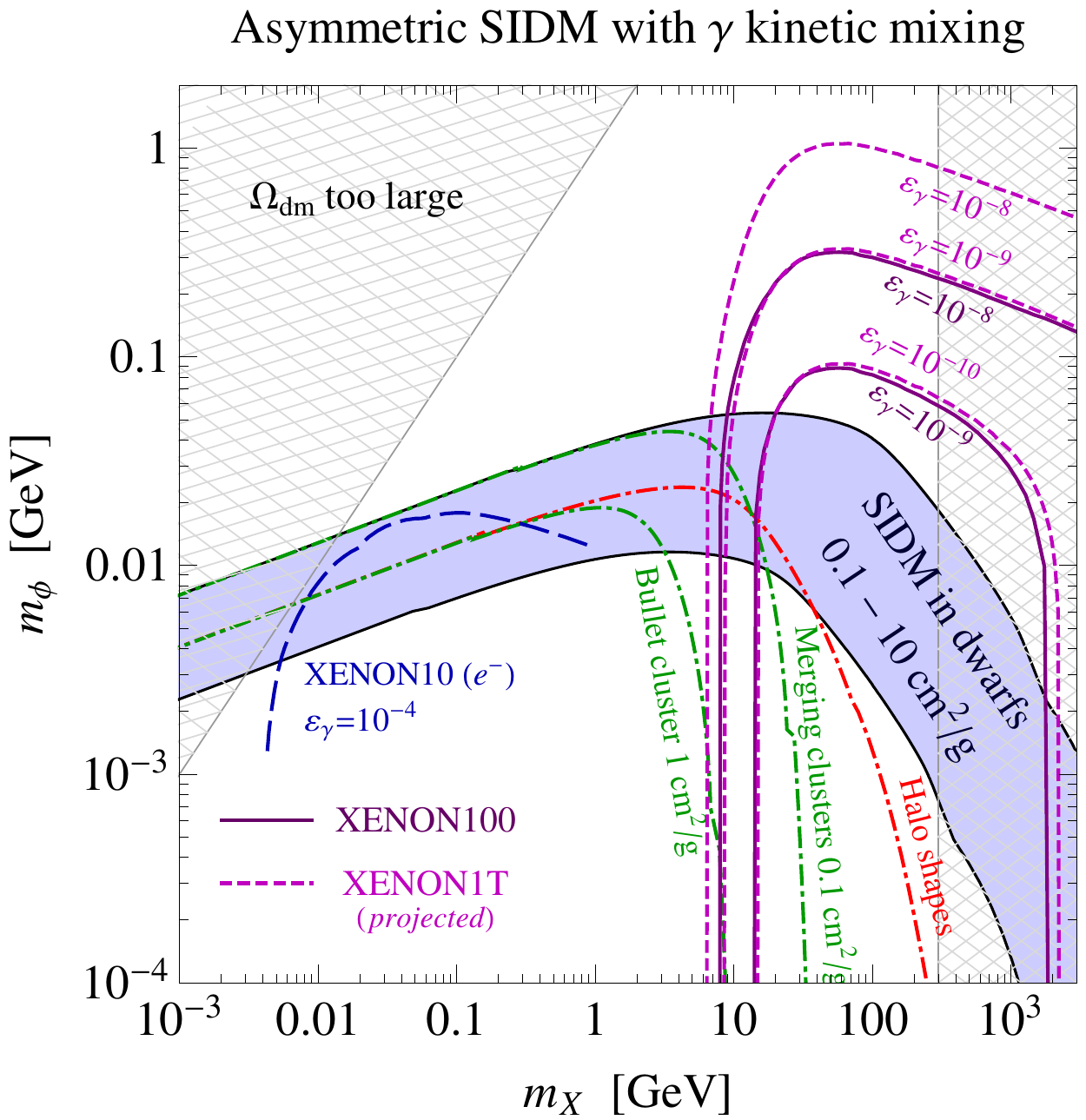} 
\includegraphics[scale=0.63]{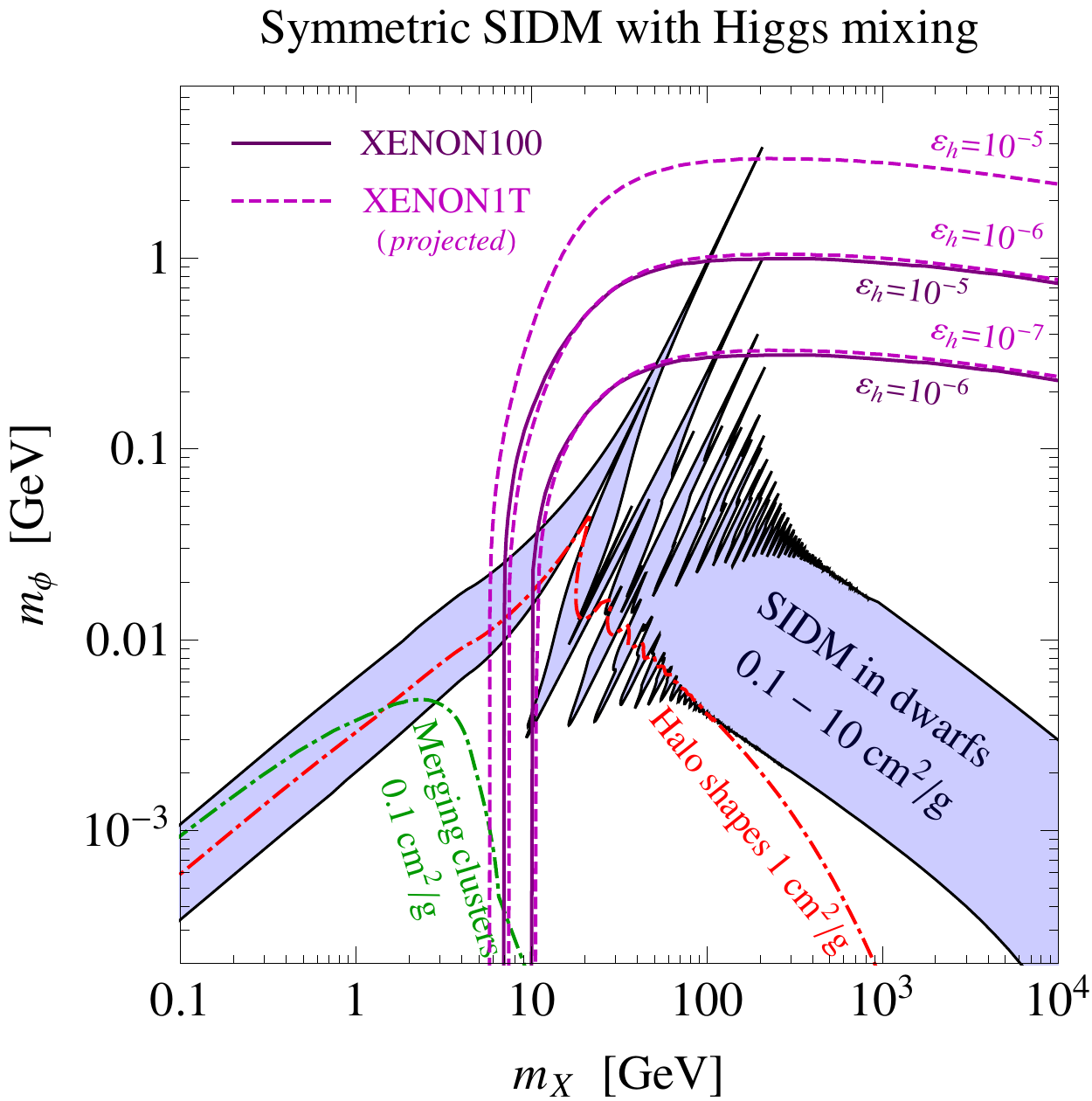}
\caption{Shaded band shows parameter space region for explaining small scale anomalies in different SIDM models.  Dot-dashed contours show constraints from astrophysical observations on larger scales (region below is excluded).  Sensitivities from XENON100 and XENON1T are shown by solid and short-dashed contours, respectively, for different $\varepsilon$ parameters.  Limits from $e^-$ recoils in XENON10 are shown by the long-dashed contour.  See text for further details.}
\label{paramspace}
\end{figure}

\begin{figure}
\includegraphics[scale=0.63]{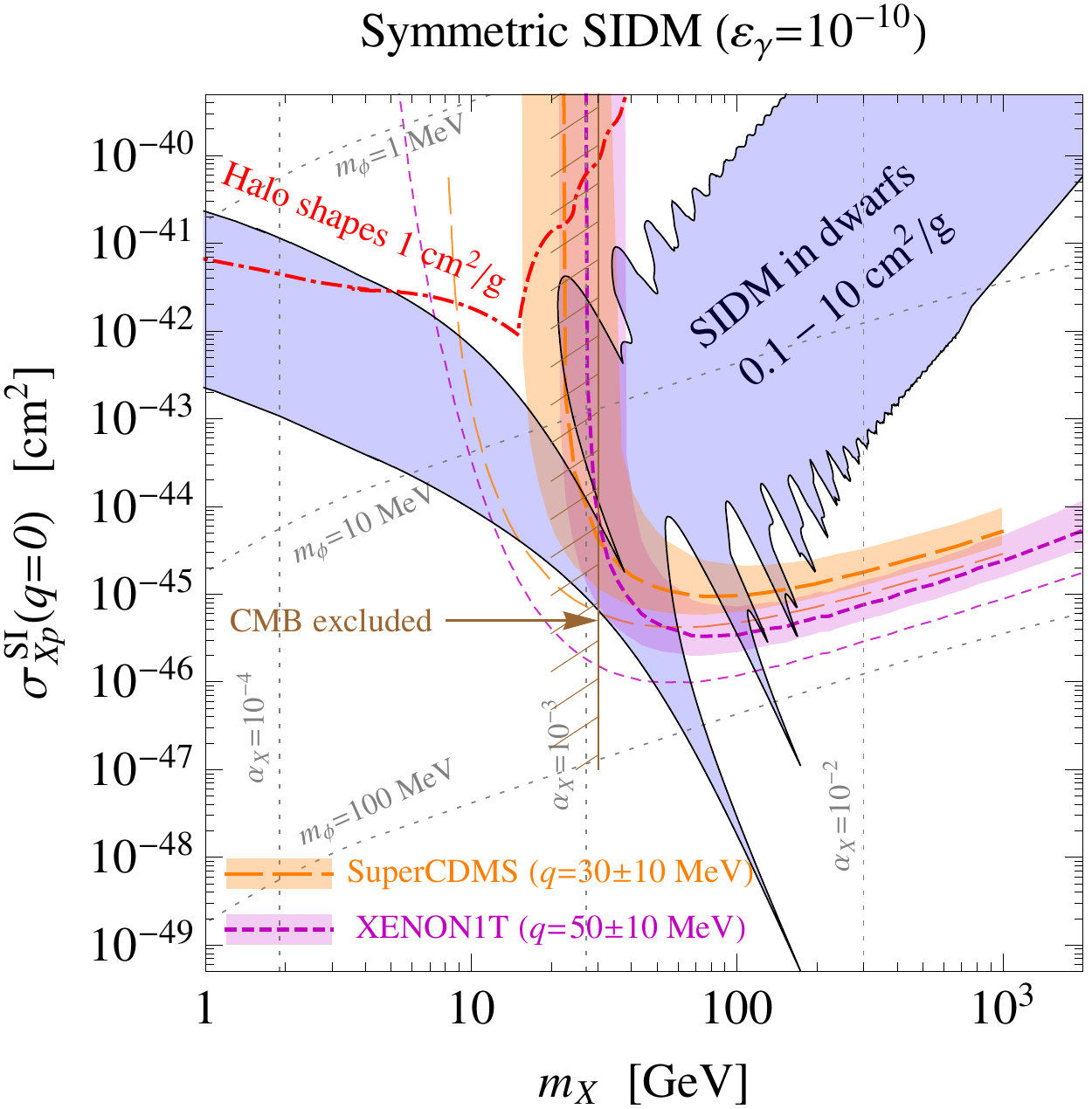} 
\includegraphics[scale=0.63]{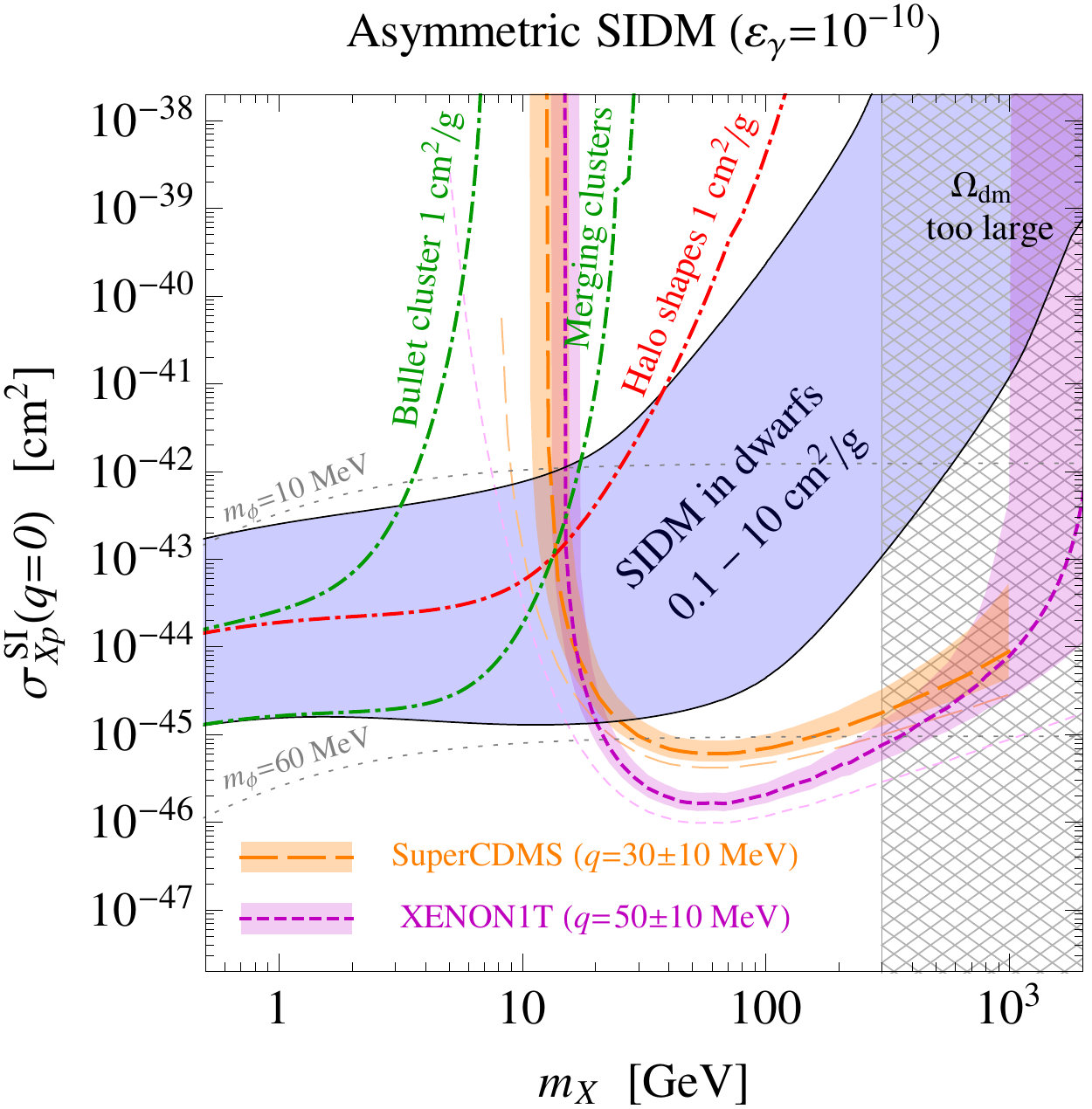}
\caption{Spin-independent direct detection cross section vs DM mass for symmetric (left) and asymmetric (right) SIDM with a vector mediator $\phi$ coupled via kinetic mixing.  Shaded band is where SIDM solves small scale structure anomalies.  Dot-dashed curves show astrophysical constraints from larger scales (excluded region lies above).  Short dashed purple curve denotes projected limit from XENON1T for $\varepsilon_\gamma = 10^{-10}$.  See text for details.}
\label{directdet}
\end{figure}

Direct detection plays a complementary role to astrophysical observables in constraining SIDM.  Our main results, presented in Figs.~\ref{paramspace} and \ref{directdet}, map out the entire parameter space for different benchmark scenarios and highlight the crucial importance of direct detection experiments, especially upcoming searches, for SIDM.  For illustrative purposes, we focus on limits from XENON100~\cite{Aprile:2012nq}, as well as projected limits for XENON1T~\cite{Aprile:2012zx}, although other experiments such as LUX~\cite{Akerib:2012ys} and SuperCDMS~\cite{Brink:2012zza} will also be important.  

In Fig.~\ref{paramspace}, we consider four benchmark SIDM models, corresponding to different scenarios with symmetric or asymmetric DM, vector or scalar mediators $\phi$, and different portals for coupling $\phi$ to the SM.  First, we summarize the different observables we consider for constraining SIDM:
\begin{itemize} 
\item Self-interactions within dwarf halos can modify the DM distribution in line with observations for $\sigma/m_X \sim 1\; {\rm cm^2/g}$~\cite{Vogelsberger:2012ku,Rocha:2012jg,Peter:2012jh,Zavala:2012us}.  The shaded bands in Figs.~\ref{paramspace} and \ref{directdet} show where these anomalies are ameliorated within a generous range, $0.1 \lesssim \sigma/m_X \lesssim 10\; {\rm cm^2/s}$.  We assume a characteristic dwarf velocity $v_0 = 30 \; {\rm km/s}$ as a compromise between smaller MW dwarf spheroidals ($v_0 \sim 10 \; {\rm km/s}$) and larger low surface brightness galaxies ($v_0 \sim 100 \; {\rm km/s}$).  

\item The ellipticity of the inner regions of DM halos, from elliptical galaxies to clusters of galaxies, suggests that $\sigma/m_X \lesssim 1 \; {\rm cm^2/s}$ on these scales~\cite{Peter:2012jh}.  Of these, elliptical galaxies provide in principle the best constraint for velocity-dependent self-interaction cross sections due to the smaller relative velocity.  Based on SIDM-only simulations (no baryons), NGC 720 seems to be inconsistent with $\sigma/m_X \gtrsim 1 \; {\rm cm^2/s}$~\cite{Peter:2012jh} and analyses of more elliptical galaxies could solidify this conclusion.  Based on this expectation, the red dot-dashed contour (labeled ``Halo shapes'') indicates the approximate upper bound $\sigma/m_X = 1 \; {\rm cm^2/g}$ from NGC 720,  adopting a characteristic velocity $v_0 = 300 \; {\rm km/s}$~\cite{Peter:2012jh}.  A similar limit on $\sigma/m_X$ for clusters  of galaxies was also suggested in Ref.~\cite{Peter:2012jh} (based on comparison of SIDM-only simulations to some LoCuSS clusters~\cite{Richard:2009yd}). 
This constraint (not shown) is weaker due to the larger relative velocity ($v_0 \sim 1000 \; {\rm km/s}$), and would lie between the halo shapes and Bullet cluster contours in Fig.~\ref{paramspace}.  It is important to emphasize that these halo shapes bounds are based on SIDM-only simulations {\it without baryons}, and we urge caution not to interpret these limits as strict constraints.  These limits (or lack thereof) depend on how much the rounded inner halos exhibited by SIDM-only simulations are modified by the aspherical baryonic component, which will be addressed in future work.

\item For the Bullet cluster constraint, we require $\sigma/m_X \lesssim 1 \; {\rm cm^2/g}$ for a relative velocity $v \approx 3000\; {\rm km/s}$~\cite{Randall:2007ph,Dawson:2012fx}.\footnote{We do not average $\sigma_T$ over a velocity distribution, but rather consider a single fixed relative velocity $v$.}  With the discoveries of other merging cluster systems, combined analyses of many such objects have the potential to improve this bound further.  We consider the potential reach of a projected constraint $\sigma/m_X \lesssim 0.1 \; {\rm cm^2/g}$ for $v \approx 2000 \; {\rm km/s}$, denoted as ``Merging clusters.''  These limits are shown by the green dot-dashed contours.

\item CMB constraints apply to symmetric DM with $s$-wave annihilation $X\bar X \to \phi\phi$, requiring $m_X \gtrsim 30 \; {\rm GeV}$ if ${\rm BR}(\phi \to e^+ e^-) \approx 1$~\cite{Lopez-Honorez:2013cua}.  This is the case for $\gamma$ kinetic mixing with $m_\phi \sim 1 - 100$ MeV, while for $Z$ mixing, the bound is weaker by a factor of ${\rm BR}(\phi \to e^+ e^-) \approx 1/7$.  The CMB exclusion region denoted by the brown hatched boundary.

\item The purple solid contours indicate constraints from XENON100~\cite{Aprile:2012nq} for different $\varepsilon$ parameters, while the purple short dashed contours indicate the projected XENON1T reach with 2.2 ton$\cdot$years exposure~\cite{Aprile:2012zx}.  
\end{itemize}
We emphasize that the quoted XENON cross section sensitivities assume an isospin-conserving contact interaction.  Here, we modify these sensitivities by the xenon proton (neutron) fraction for the case of $\gamma$ kinetic ($Z$) mixing.  We also include a $q^2$-dependent form factor, described in Eq.~\eqref{formfactor}, since the typical momentum transfer $q$ is comparable to $m_\phi$.  We take $q \approx 50$ MeV for xenon.

The first benchmark model is the case of symmetric SIDM with a vector mediator $\phi$ that couples through $\gamma$ kinetic mixing.  The constraints on this model are shown in Fig.~\ref{paramspace} (upper left) as a function of $(m_X,m_\phi)$.  The DM relic density, determined by thermal freeze out, fixes $\alpha_X \approx 4 \times 10^{-5} \times (m_X/{\rm GeV})$. Self-interactions are a combination of attractive and repulsive, and the peak-like features correspond to where scattering is resonantly enhanced.  The halo shapes bound (with excluded regions below the red dot-dashed contour, when taken at face value) is weak due to the velocity-dependence of the self-interaction cross section.  However, the SIDM region below $30$ GeV is excluded by CMB.  The sensitivity contours for XENON100 and XENON1T are shown for different values of $\varepsilon_\gamma$, excluding SIDM parameter space below the curves.

The second benchmark model shown in Fig.~\ref{paramspace} (upper right) is the same as the previous case, but with $\phi$ coupled through $Z$ mixing.  The self-interaction observables and relic density are unchanged.  However, the CMB bound is weakened since $\phi$ decays predominantly to neutrinos.  Direct detection sensitivities are also affected due to the different couplings, but are quantitatively similar to the kinetic mixing scenario.

Third, we consider a scenario with asymmetric SIDM, shown in Fig.~\ref{paramspace} (lower left), with fixed $\alpha_X=10^{-2}$.  For a mediator, we consider a vector $\phi$ coupled through $\gamma$ kinetic mixing.  No resonant features are present since scattering is purely repulsive (no bound states).  Also, since annihilation does not occur, indirect detection signals are absent and the CMB bound does not apply.  Thus, we extend the parameter range to sub-GeV DM masses.    The dot-dashed contours show how constraints from astrophysical observations can constrain the low mass regime (below the curves is excluded).  Outside the range $m_\phi < m_X \lesssim 300$ GeV there is insufficient annihilation of the symmetric density (hatched region), although larger values of $\alpha_X$ allow for larger $m_X$.\footnote{We neglect a narrow parameter strip near $m_\phi \approx 2 m_X$ where resonant $s$-channel annihilation can occur.}  In addition to the XENON100 and XENON1T contours, we also show the reach for direct detection via electron recoils for constraining sub-GeV SIDM.  The current limits from XENON10 are shown by the long dashed curve for $\varepsilon_\gamma = 10^{-4}$.  (For such large values of $\varepsilon_\gamma$, the $m_X \gtrsim 1$ GeV region is strongly exluded by direct detection via nuclear recoils~\cite{Essig:2011nj}.)

Our final benchmark model is symmetric SIDM with a scalar mediator coupled via Higgs mixing, shown in Fig.~\ref{paramspace} (lower right).  The DM relic density fixes $\alpha_X \approx 10^{-4} \times  (m_X/{\rm GeV})$. Self-interactions are purely attractive.  Indirect detection bounds do not apply since annihilation is $p$-wave suppressed.  
DM self-scattering is purely attractive, and large resonant features are evident.  Since annihilation is $p$-wave, the CMB bound does not apply.  

We make several comments on the results shown in Fig.~\ref{paramspace}:
\begin{itemize}
\item The shaded bands show that small scale structure anomalies can be explained for a wide range of SIDM mass, from sub-GeV to multi-TeV, and the mediator mass is typically in the $\sim 1 - 100$ MeV range.
\item Direct detection limits constrain SIDM above $\sim 10$ GeV; the exclusion region lies below the solid (dashed) purple curves for XENON100 (XENON1T).  In this regime, self-interactions are velocity-dependent, and SIDM easily evades the Bullet cluster and halo shapes bounds from larger scales.
\item For DM below 10 GeV, astrophysical constraints from indirect detection and structure observables are most important for constraining SIDM.  In this regime, the self-interaction cross section is more velocity-independent, and future merging cluster studies will be important.  Direct detection via electron recoils provides another avenue to explore low mass SIDM, albeit with larger $\varepsilon_\gamma$.  Direct detection via nuclear recoils is below threshold and provides no constraint.
\item For vector mediator scenarios, ton-scale direct detection experiments will cover almost the entire SIDM parameter space for $m_X \gtrsim 25$ GeV, down to the lower bound $\varepsilon_{\gamma,Z} \sim 10^{-10}$ imposed by BBN.  Current experiments have no constraint for $\varepsilon_{\gamma,Z} \sim 10^{-10}$.
\item For a scalar mediator with Higgs mixing, this scenario is almost completely excluded by XENON100 unless $m_X \lesssim 5$ GeV.  This case is more strongly constrained by direct detection compared the vector mediator case since $\phi$ couplings to SM fermions are proportional to mass.  Therefore, the effective $\phi$-nucleon coupling entering direction detection is enhanced compared to the $\phi$-electron coupling constrained by BBN.
\end{itemize}

In Fig.~\ref{directdet}, we show how SIDM benchmark models map onto the direct detection plane of cross section vs mass.  For simplicity, we focus on the case where the mediator $\phi$ is a vector.  We assume kinetic mixing with $\varepsilon_\gamma=10^{-10}$ motivated by BBN constraints.  The different panels are for symmetric DM (left) or asymmetric DM (right), corresponding to the first two benchmarks shown in Fig.~\ref{paramspace} (top left and right).  The vertical axis is the spin-independent cross section for scattering on protons in the $q^2 = 0$ limit.  We emphasize that this cross section, however, does not appear in the direct detection rate by itself but must be multiplied by a form factor taking into account the finite $q^2$.  The short-dashed purple curve shows the sensitivity reach of XENON1T taking $q = 50$ MeV, while the long-dashed orange curve shows the sensitivity reach of SuperCDMS (SNOLAB) based on $385 \; {\rm kg}\cdot{\rm years}$ exposure~\cite{Sander:2012nia} taking $q=30$ MeV.  To illustrate the importance of the $q^2$-dependent form factor, the surrounding band shows how the XENON1T and CDMS limits change by varying $q$ in the range $\pm 10$ MeV, while the corresponding thin countours shows how the limits would appear for $q=0$.  As above, the shaded band shows where SIDM solves small scale anomalies, while the dot-dashed contours show astrophysical constraints on self-interactions on larger scales.  The hatched boundary shows the CMB exclusion limit on symmetric DM.


\section{Conclusions}
\label{sec:conclusions}

SIDM is a simple and well-motivated scenario that can explain small scale structure anomalies observed in dwarf galaxies.  Self-interactions arise through a $1-100$ MeV mediator particle in the dark sector.  Requiring that mediator particles decay to SM states before BBN implies a minimal coupling between DM and the SM, which can be probed in direct detection experiments.  In this work, we have considered three mechanisms for coupling the mediator to the SM and studied the direct detection implications of SIDM. With a light mediator favored by the small scale structure of the Universe, SIDM has direct detection features different from usual WIMPs. Current and future direct detection experiments are sensitive to SIDM candidates, even if the coupling between two sectors is extremely feeble. SIDM also interacts with target nuclei with momentum-dependent interactions, which may provide a mechanism to reconcile some of the recent dark matter direct detection hints~\cite{Chang:2009yt,Fornengo:2011sz,Foot:2013msa}. For $\gamma$ or $Z$ mixing, we found that current direct detection limits provide no constraint on SIDM if the coupling strength saturates the lower bound from the cosmological considerations. However, future ton-scale experiments will explore the entire parameter range for SIDM above $\sim 20$ GeV. On the other hand, for Higgs mixing, the same mass range is already excluded by XENON100. Thus, astrophysical observations and direct detection experiments complement each other in the search for SIDM candidates.

\vspace{0.5cm}

{\it Acknowledgments}: We thank Sunil Golwala, Moira Gresham, Rafael Lang, David Sanford, and Andrew Zentner for useful discussions, as well as the organizers of the Harvard SIDM 2013 workshop for a stimulating meeting and hospitality during the completion of this work.  MK is supported by NSF Grant No.~PHY-1214648. HBY is supported by startup funds from the UCR.  ST is supported by the DOE under contract DE-SC0007859 and NASA Astrophysics Theory Grant NNX11AI17G.  ST would also like to thank Aspen Center for Physics and NSF Grant No.~1066293, as well as CETUP* (Center for Theoretical Underground Physics and Related Areas), supported by DOE Grant No. DE-SC0010137 and NSF Grant No. PHY-1342611, for hospitality and partial support during the completion of this work.

\bibliography{ddbib}

\end{document}